# The Validity of Company Valuation
# Using Discounted Cash Flow Methods

## Florian Steiger[1]

### Seminar Paper
#### Fall 2008

## Abstract

This paper closely examines theoretical and practical aspects of the widely used discounted cash flows (DCF) valuation method. It assesses its potentials as well as several weaknesses. A special emphasize is being put on the valuation of companies using the DCF method. The paper finds that the discounted cash flow method is a powerful tool to analyze even complex situations. However, the DCF method is subject to massive assumption bias and even slight changes in the underlying assumptions of an analysis can drastically alter the valuation results. A practical example of these implications is given using a scenario analysis.

_______________
[1] Author: Florian Steiger, European Business School, e-mail: florian.steiger@post.harvard.edu

**Table of Contents**

**List of abbreviations**

| | |
|---|---|
| APV | Adjusted Present Value |
| bp | Base Point (equal to 0.01%) |
| Capex | Capital Expenditure |
| CAGR | Compounded Annual Growth Rate |
| CAPM | Capital Asset Pricing Model |
| COD | Cost of Debt |
| COE | Cost of Equity |
| D&A | Depreciation and Amortization |
| DCF | Discounted Cash Flow |
| EBIT | Earnings Before Interests and Taxes |
| EBITDA | Earnings Before Interests, Taxes, Depreciation and Amortization |
| EURm | Millions of Euro |
| EV | Enterprise Value |
| Eq. V. | Equity Value |
| FCF | Free Cash Flow |
| FCFE | Free Cash Flow to Equity |
| FCFF | Free Cash Flow to Firm |
| IPO | Initial Public Offering |
| LBO | Leveraged Buyout |
| LIBOR | London Interbank Offer Rate |
| M&A | Mergers and Acquisitions |
| NI | Net Income |
| NOPAT | Net Operating Profit After Taxes |
| NPV | Net Present Value |
| P / E | Price Earnings Ratio |
| r | Discount Rate |
| ROA | Return on Assets |
| ROE | Return on Equity |
| SIC | Standard Industry Classification |
| t or T | Tax Rate |
| T-Bill | US Treasury Bill |
| T-Bond | US Treasury Bond |
| TV | Terminal Value |

# List of figures and tables

# 1     Introduction

The goal of this paper is to introduce the reader to the method of company valuation using discounted cash flows, often referred to as "DCF". The DCF method is a standard procedure in modern finance and it is therefore very important to thoroughly understand how the method works and what its limitations and their implications are. Although this paper is on a basic level, it requires some knowledge of accounting and corporate finance, as well as a good understanding of general economic coherencies, since not every topic can be explained in detail due to size limitations.

## 1.1     Problem Definition and Objective

Since the beginning of the year 2008, Goldman Sachs has advised clients on merger and acquisition (M&A) deals with aggregated enterprise values (EV) of more than EURm 475,000 according to recent league tables (Thomson One Banker, 2008). There are "probably almost as many motives for M&As as there are bidder and targets" (Mukherjee, Kiymaz, & Bake, 2004, p. 8), but the transaction volumes indicate the importance that M&A activities have for the worldwide economy and underline the necessity for efficient methods to adequately value companies.

The DCF method is based upon forward looking data and therefore requires a relatively large amount of predictions for the future business situation of the company and the economy in general. Minor changes in the underlying assumptions will result in large differences in the company's value. It is therefore very important to know which assumptions are used and how they influence the outcome of the analysis. For this reason, this paper will introduce the key input factors that are needed for the DCF analysis and examine the consequences that changes in the assumptions have on the company value.

The DCF analysis is a very powerful tool that is not only used to value companies but also to price initial public offerings (IPOs) and other financial assets. It is such a powerful tool in finance, that it is so widely used by professionals in investment banks, consultancies and managers around the world for a range of tasks that it is even referred to as "the heart of most corporate capital-budgeting systems" (Luehrman, 1998, p. 51).

## 1.2    Course of the Investigation

This paper begins with a brief introduction to valuation techniques in general and shows how valuation techniques can be used to assess a company's value. Afterwards the basic idea behind the DCF valuation technique will be introduced and the key input factors will be explained and discussed, since it is most important to gain a deep understanding on how the input is computed to state the company value. In the next step a sensitivity analysis will be conducted using BASF as an example to explain how varying input will lead to different results. In the end, a conclusion will be drawn on the benefits and shortfalls of the DCF valuation technique.

## 2    Company valuation

### 2.1    General Goal and Use of Company Valuation

The goal of company valuation is to give owners, potential buyers and other interested stakeholders an approximate value of what a company is worth. There are different approaches to determine this value but some general guidelines apply to all of them.

In general there are two kind of possible takeover approaches. An interested buyer could either buy the assets of a company, known as asset deal, or the buyer could take over a majority of the company's equity, known as share deal.[2] Since taking over the assets will not transfer ownership of the legal entity known as "the company", share deals are much more common in large transactions. Due to the financing of a company by debt and equity, valuation techniques that focus on share deals either value the equity, resulting in the *equity value* (Eq. V.) or the total liabilities, stating the *enterprise value* (EV) or *firm value* (FV). It is possible to derive the EV from the Eq. V. and vice versa (Bodie, Kane, & Marcus, 2008, pp. 630-631) by using the following formula:

$$EV - Net\ Debt - Corporate\ Adjustments = Eq.V.$$

Net debt and the corporate adjustments are derived with the following definitions:

$$Net\ Debt = Long\ term\ debt + Short\ term\ debt + Capitalized\ leases$$
$$+ Other\ interest\ carrying\ liabilites - Cash\ and\ cashlike\ assets$$

---

[2]Actually there are more possibilities to gain ownership of a company, like a debt-to-equity swap, where debt holders offer the equity holders to swap their debt into equity of the company and therewith gain equity ownership. This usually happens with companies that are in financial distress like insolvency or bankruptcy.

$$Corporate\ Adjustments$$
$$= Minority\ interests + PV\ of\ pension\ deficit$$
$$+ Offbalance\ sheet\ obligations \pm Associated\ companies$$

## 2.2    Other Valuation Methods

There are many other valuation techniques besides the DCF approach which are commonly used. In fact, most of the time various techniques are used and the results are then compared to each other to increase the confidence that the result is reasonable.

A widely used method is the so-called *trading comparables* analysis. In this method a peer group of listed companies is built, usually using firms with similar standard industry classification (SIC) and other similarities to the target company like geographic focus, financing structure, and client segments. If the company is listed, the equity value is simply the market capitalization[3]. The EV can be calculated based on this Eq. V. as described above. Then some multiples are calculated to state relationship between EV and Eq. V. to a company's fundamental data. Usually the multiples are the following:

$$\frac{EV}{Sales} \qquad \frac{EV}{EBIDA} \qquad \frac{EV}{EBIT} \qquad \frac{Eq.V.}{Net\ Income}[4]$$

The median and arithmetic average of these multiples is then calculated for the peer group.[5] These figures are a good approximation for a target's EV and Eq. V., but they tend to be lower than actual transaction values, since trading comparables do not include majority premiums that have to be paid when acquiring a majority stake in a company.

A similar approach to the trading comparables method is the transaction comparables valuation approach. It uses the same multiples, but the peer group consists of previous transactions and therefore includes all premiums that arise during transactions. This method is very reliable but since it is very difficult to find previous transactions that are similar, it is difficult to build peer groups that are statistically significant[6]. These two methods, in combination with the DCF are the most widely used in modern finance.

---

[3] $Market\ Cap = Share\ Price * Number\ of\ shares\ outstanding$
[4] The $\frac{Eq.V.}{Net\ Income}$ is the same as the trailing (historical) $\frac{P}{E}$ ratio
[5] Please see table 2 in the appendix for an exemplary trading comparables analysis of the European car rental market
[6] Please see table 3 in the appendix for an example

## 3    The Discounted Cash Flow Valuation Method

### 3.1    Approach of the Discounted Cash Flow Valuation

The DCF method values the company on basis of the *net present value* (NPV) of its future free cash flows which are discounted by an appropriate discount rate. The formula for determining the NPV of numerous future cash flows is shown below. It can be found in various sources, e.g. in "Financial Management – Theory and Practice" (Brigham & Gapenski, 1997, p. 254).

$$NPV = \sum_{t=0}^{n} \frac{FCF_t}{(1+r)^t}$$

The free cash flow is the amount of "cash not required for operations or reinvestment" (Brealey, Myers, & Allen, 2006, p. 998). Another possibility to analyze a company's value using discounted cash flows is the *adjusted present value* (APV). The APV is the net present value of the company's free cash flows assuming pure equity financing and adding the present value of any financing side effect, like tax shield (Brealey, Myers, & Allen, 2006, p. 993) In general you can say, that the APV is based on the "principle of value additivity" (Luehrmann, 1997, S. 135). However, APV and NPV lead to the same result.

Since the DCF method is a valuation technique that is based on predictions, a scenario analysis is usually conducted to examine the effects of changes in the underlying assumptions. Such a scenario analysis is usually based on three scenarios, namely the "base case" or "management scenario" that uses the management's estimations for the relevant metrics, a "bull case" which uses very optimistic assumptions and a "bear case" that calculates the company's value if it performs badly.

The process of valuing a company with the DCF method contains different stages. In the first stage scenarios are developed to predict future free cash flows (FCF) for the next five to ten years. Afterwards, an appropriate discount rate, the *weighted average cost of capital* (WACC) has to be determined to discount all future FCFs to calculate their NPVs. In the next step the *terminal value* (TV) has to be identified. The TV is the net present value of all future cash flows that accrue after the time period that is covered by the scenario analysis. In the last step the net present values of the cash flows are summed up with the terminal value.[7]

$$Company\ Value = \sum_{t=0}^{n} \frac{FCF_t}{(1+r)^t} + Terminal\ Value$$

## 3.2    Calculation of the Free Cash Flow

### 3.2.1 Cash Flow to Firm and Cash Flow to Equity

There are two ways of using cash flows for the DCF valuation. You can either use the *free cash flow to the firm* (FCFF) which is the cash flow that is available to debt- and equity holders, or you can use *the free cash flow to equity* (FCFE) which is the cash flow that is available to the company's equity holders only.

When using the FCFF, all inputs have to be based on accounting figures that are calculated before any interest payments are paid out to the debt holders. The FCFE in contrast uses figures from which interest payments have already been deducted. Using the FCFF as base for the analysis will result in the enterprise value of the company, using the FCFE will give the equity value. Since an acquirer usually takes over all liabilities, debt and equity, the FCFF is more relevant than the equity approach.

The FCFF is calculated by deducting taxes from the company's earnings before interest and taxes (EBIT), resulting in the net operating profit after tax (NOPAT). All calculatory costs (e.g. D&A) are then added back, since they do not express any cash flows. The capital expenditure (Capex) is deducted. It is a cash outflow that is not reflected in the income statement, because Capex is activated on the asset side of the balance sheet. The increase in net working capital (NWC) is also deducted, because it is does not represent any actual cash flows. The formula for calculating the FCFF is shown below. (Damodaran, 1996, p. 237)

$$FCFF = NOPAT + D\&A - Capex - Increase\ in\ NWC$$

There are more methods that can be used to calculate the FCFF, but they will all result in the same value.

---

[7]$Terminal\ Value = \sum_{t=n+1}^{\infty} \frac{FCF_t}{(1+r)^t}$

### 3.2.2 Building Future Scenarios

Deriving the NPV of the free cash flows that accrue in the scenario period is very complex, because all these cash flows are based on assumptions. The method therefore requires a detailed picture of the company's future situation, e.g. EBIT and Capex. Predictions are usually made for the next five to fifteen years. The NPV of the cash flows accruing after this scenario period is included in the terminal value, which is derived using much less assumptions. These predictions are usually based on historical data, but may also reflect changes in the company's business plan, industry or in the global economy.

To provide a detailed view on how the company's value might be affected by a change in the underlying assumption, a scenario analysis is usually conducted. In the bear case scenario, low assumptions for rates of growth and margins are used to build a very pessimistic scenario. In the bull case the opposite is the case, all assumptions are very optimistic. These two cases mark the boundaries of where in between the fair value of the company should be with a high certainty. Of course, additional scenario and risk testing methods like value at risk using a Monte Carlo Simulation can be used to further evaluate any risks.

The most important scenario in the valuation of a company is the base case. In this case the management's predictions and opinions regarding the future development of the company, its relevant markets and competitors are used to build the scenario that is most likely to happen. However, attention has to be paid to the reliability of any management provided figures, since managers often have a personal incentive to increase the takeover price and therefore might provide biased estimates.

Another item that is usually included are potential synergies between the target and the acquirer. If the potential acquirer is a strategic acquirer who runs a similar business, many synergies can be realized. This will allow the strategic bidder to offer a higher price than a financial bidder, like a private equity funds for example.

### 3.3    The Weighted Average Cost of Capital

Determining the discount rate requires extensive analysis of the company's financing structure and the current market conditions. The rate that is used to discount the FCFs is called the *weighted average cost of capital* (WACC). The WACC is one of the most important input factors in the DCF model. Small changes in the WACC will cause large changes in the firm value. The WACC is calculated by weighting the sources of capital according to the company's financial structure and then multiplying them with their costs. Therefore the formula for the WACC calculation is:[8]

$$WACC = \frac{Equity}{Debt + Equity} * Cost\ of\ Equity + \frac{Debt}{Debt + Equity} * Cost\ of\ Debt$$

### 3.3.1 Cost of Equity

The *cost of equity* (COE) is calculated with the help of the *capital asset pricing model* (CAPM). The CAPM reveals the return that investors require for bearing the risk of holding a company's share. This required return is the *return on equity* (ROE) that investors demand to bear the risk of holding the company's share, and is therefore equivalent to the company's cost of equity. According to the CAPM, the required ROE, or in this case the COE is derived with the following formula (Ross, Westerfield, & Jordan, 2008, p. 426):

$$COE = r_f + \beta(r_m - r_f)$$

Although the risk-free interest rate is the yield on T-Bills or T-Bonds, professionals use the *London Interbank Offer Rates* (LIBOR) as an approximation for the short-term risk-free interest rates, since ". . . treasury rates are too low to be used as risk-free rates . . . " (Hull, 2008, p. 74) It is therefore common to use the LIBOR as the risk-free rate for valuation purposes.

The input factor $\beta$ is the risk, that holding the stock will add to the investor's portfolio[9] (Rhaiem, Ben, & Mabrouk, 2007, p. 80). It is derived using linear regression analysis, where the excess return of the stock is the dependent variable and the excess market return is the independent variable. The beta is the slope of the regression line. (Brealey, Myers, & Allen, 2006, p. 220) Beta is an empirical determined input factor that is also based on the company's historical level of leverage, because higher leverage ratios increase the shareholder's risk. Since the company's level of leverage often changes during a transaction, the beta has to be adjusted for this change by unlevering and relevering to the new capital structure. If the company is not listed there is no data

---

[8]In case of any preferred share outstanding, the formula has to be rearranged to include this source of financing as well. The adjusted formula will be as following:

$WACC = \frac{Equity}{Assets} * COE + \frac{Debt}{Assets} * COD + \frac{Preferred}{Assets} * Cost\ of\ Preferred\ Capital$

[9] $\beta = \frac{cov(Stock, Market)}{var(Market)}$

available to compute a linear regression. As a consequence, a peer group of similar companies is set up and the median of their unlevered betas is then relevered to fit the target's financing structure. Although the CAPM approach is very useful to estimate the cost of equity, some scientists argue that the CAPM was developed for liquid assets (Michailetz, Artemenkov, & Artemenkov, 2007, p. 44), and therefore its significance for the valuation of illiquid assets, like non-listed companies should be subject to further research.

### 3.3.2 Cost of Debt

The *cost of debt* (COD) is the interest rate that a company has to pay on its outstanding debt. The most influencing factor on the COD is a company's credit rating. A company with an investment grade credit rating [10] (e.g.: S&P AAA) is able to borrow at considerably lower interest rates than a company that is rated as non-investment grade (e.g.: S&P BB-). The difference between the risk-free interest rate and the interest rate that a company has to pay to borrow money is called the company's credit spread. The credit spread does not only depend on a company's credit worthiness, but is also determined by market conditions. An indicator for these conditions is the spread of the USD 3m LIBOR vs. the 3m T-Bills[11] depicted in figure 1 in the appendix (Bloomberg Professional Database, 2008). The chart reflects a massive widening in credit spreads that occurred in August 2007 after numerous banks and hedge funds announced a massive exposure to the so-called subprime mortgage market. The dependence of overall market conditions should be kept in mind when calculating the COD. Especially when the company has a high leverage ratio, special attention has to be paid to the credit markets.

Interest rate costs are tax deductible in most economies, so that the true COD is lower than the interest rate a company pays out to its debt holders[12]. Due to the fact that taxation laws are very different around the world, a very thorough analysis is needed to verify how much of the interest costs are deductible. The COD after tax can be calculated as following, where $i$ is the interest rate on outstanding debt and $t$ is the effective tax rate paid by the company:

---

[10] Please see table 1 for an overview of long term credit rating scales of different rating agencies

[11] Another widely used benchmark to assess the credit spread is the iTraxx Europe index, a credit index consisting of 125 investment grade companies in Europe

[12] Assuming the fact that the company is paying taxes from which the COD can be deducted

$$COD = i * (1 - t)$$

If the company has different kinds of debt outstanding, the COD is the weighted average cost of debt of these different tranches, adjusted for tax:[13]

$$COD = (1 - t) * \sum_{a=1}^{n} w_a i_a$$

### 3.3.3 Summary

By plugging in the formulas for the COE and COD, we get the full formula for the WACC including all factors that influence the discount rate:

$$WACC = \frac{E}{D+E} * \left(r_f + \beta\left(r_m - r_f\right)\right) + \frac{D}{D+E} * i * (1 - t)$$

The WACC is therefore determined by the COE, which is derived by applying the CAPM with its underlying assumptions for beta. The COD is derived from the interest rate that the company has to pay to its debt holders and by the tax rate that the corporation has to pay on its profits. Changing the assumptions for the cost of capital will have large effects on the result of the overall valuation process.

The WACC of a company is dependent on a variety of economic factors. Especially the company's industry and the steadiness of its cash flows influence it. Companies with stable cash flows in mature industries with low growth rates will typically have low capital costs (Morningstar, 2007, pp. 1-2). For example, Bayer will have a substantially lower WACC than Conergy.

The WACC is used to discount the FCFs that we predicted in our scenario analysis. The result is the NPV of the company in the scenario period, to which we will later add the terminal value, which also makes uses of the WACC.

Using current figures for beta, risk-free rate, credit spread, and interest costs will lead to a fairly realistic approximation for the discount rate in most cases. However, to get an exact value, the company's future WACC must be used. Therefore, all input factors of the WACC formula have to be predicted, resulting in leeway for the outcome of the DCF analysis.

---

[13]The weights are calculated by dividing the market value of a tranche by the market value of total debt outstanding: $w_a = \frac{Market\ value\ of\ tranche}{Market\ value\ of\ total\ debt}$

### 3.4    Calculation of the Terminal Value

The terminal value is the NPV of all future cash flows that accrue after the time period that is covered by the scenario analysis. Due to the fact that it is very difficult to estimate precise figures showing how a company will develop over a long period of time, the terminal value is based on average growth expectations, which are easier to predict.

The idea behind the terminal value is to assume constant growth rates for the time following the time period that was analyzed more extensively. The constant perpetual growth rate $g$, together with the WACC as the discount rate $r$ allows for the use of a simple dividend discount model to determine the terminal value. Therefore the TV can be expresses as[14] (Beranek & Howe, 1990, p. 193), where the FCF is one period before the TV period:

$$TV = \sum_{n=1}^{\infty} \frac{FCF_{TV} * (1 + g)^n}{(1 + r)^n} = \frac{FCF_{TV}(1 + g)}{r - g}$$

Since all these cash flows are discounted to a date in the future, the TV has to be discounted again to give us the NPV of all free cash flows that occur after the scenario predicted period.

The determination of the perpetual growth rate is one of the most important and complex tasks of the whole DCF analysis process, since minor changes in this rate will have major effects on the TV and therefore on the firm value in total. The huge range of values that result from a change in this growth rate will be examined in a case study later on in this chapter. In most cases a perpetual growth rate should be between 0% and 5%. It has to be positive since in the long-term, the economy is always growing. However, according to economists, any growth rate above 5% is not sustainable on the long-term. The perpetual growth rate should be in line with the nominal GDP growth. (JP Morgan Chase, 2006).

Due to the fact, that the TV often accounts for more than half of the total company value, special attention has to be paid to its calculation and input coefficients. As discussed in the case study later in this paper, even very small changes that might not even be significant from an economist's perspective will result in substantial changes in the company value. Therefore it is very easy to move the TV into the desired direction without having to drastically change any underlying business predictions, like EBIT margin or capital expenses.

---

[14] $TV = \frac{FCF*(1+g)^1}{(1+r)^1} + \frac{FCF*(1+g)^2}{(1+r)^2} + \frac{FCF*(1+g)^3}{(1+r)^3} ... \frac{FCF*(1+g)^n}{(1+r)^n}$ which can be mathematically rearranged to equal the formula given in the text

### 3.5     Determination of Company Value

After having determined the NPV of the cash flows accruing within the scenario period and the TV, the TV is discounted to its NPV. Both NPVs are then added together to give the enterprise value or the equity value, depending on whether the valuation is based on FCFFs or FCFEs:

$$Company\ Value = \sum_{t=0}^{n} \frac{FCF_t}{(1+r)^t} + \frac{TV}{(1+r)^{n+1}}$$

Usually the company value is calculated using different levels of leverage to find an optimal financing structure. The determined company value can then be used for further analysis, e.g. the equity value could be divided by the number of shares outstanding to determine a fair share price for listed companies.

## 4     Validity of the Discounted Cash Flow Valuation Approach

### 4.1     Case Study: BASF

To demonstrate the wide range of possible results of the DCF analysis, this paper will now analyze the BASF stock and the DCF's sensitivity to changes in the WACC, the perpetual growth rate, and sales growth. For this purpose, a base scenario based on broker estimates (Credit Suisse Equity Research, 2008) will be built to obtain a fair reference value for one BASF stock. Afterwards a sensitivity analysis will be conducted to examine the effects on this reference price that modifying factors will have.

The base case scenario uses the estimates by Credit Suisse analysts for the cash flow forecasts for the years 2008 to 2013. The unlevered beta was determined to be 0.9 using a linear regression model leading to the cost of equity of 10.3%. BASF's current credit rating results in a credit spread of 500bp according to analysts (Credit Suisse Equity Research, 2008). This leads to a WACC of 9.0%. Furthermore we assume the perpetual growth rate to be equal to 1.5%. Discounting the predicted free cash flows to the firm for the years 2008 to 2013 using the WACC of 9.0% and then adding the discounted terminal value results in an enterprise value of EURm 67,850. Please see tables 3 - 7 for the exact calculations.

| Period | 2008E | 2009E | 2010E | 2011E | 2012E | 2013E | TV |
|--------|-------|-------|-------|-------|-------|-------|------|
| FCFF | 4,284 | 4,405 | 4,866 | 5,409 | 6,148 | 6,212 | - |
| NPV | 3,930 | 3,708 | 3,758 | 3,832 | 3,996 | 3,704 | 44,923 |
| EV | 67,850 | | | | | | |

Table 4: Case Study: Calculation of the enterprise value

It is remarkable that the terminal value accounts for EURm 44,923 of the total EV. This makes obvious, that the outcome of the DCF analysis is highly sensitive to changes in the perpetual growth rate, since it has a major effect on the TV. Having determined the EV, net debt and corporate adjustments are deducted from the EV to calculate the equity value of EURm 55,332. The equity value is then divided by the number of shares outstanding. The result of EUR 58.49 is the fair price for one BASF share given the underlying assumptions. Knowing the fact that the current share price equals only EUR 39.41 (Thomson Reuters, 2008), this would make the BASF share a great investment if you believe that the underlying assumptions are valid. This share price will serve as the reference value for the sensitivity analysis, since it lies in between of most research analyst's target price for BASF.

## 4.2    Sensitivity Analysis

To investigate the sensitivity of the DCF method, the BASF case study developed above will be used. The changes that occur in the share price will be stated as percentage offset from the base case share price of EUR 58.49.

The WACC and the perpetual growth rate are two main input factors that have large effect on the outcome of the analysis. Therefore the table below shows the result of the sensitivity analysis regarding those two factors. The base case assumptions of 9.0% for the WACC and 1.5% for the perpetual growth rate are highlighted in dark blue.

| | | WACC (%) | | | | | | | | |
|---|---|---|---|---|---|---|---|---|---|---|
| | | 7.0% | 7.5% | 8.0% | 8.5% | 9.0% | 9.5% | 10.0% | 10.5% | 11.0% |
| **Perpetual growth rate (%)** | 0.0% | 19.2% | 9.0% | 0.2% | -7.6% | -14.5% | -20.7% | -26.2% | -31.2% | -31.2% |
| | 0.5% | 27.2% | 15.8% | 5.9% | -2.7% | -10.3% | -17.0% | -23.0% | -28.3% | -28.3% |
| | 1.0% | 36.6% | 23.6% | 12.5% | 2.9% | -5.4% | -12.8% | -19.3% | -25.1% | -25.1% |
| | 1.5% | 47.6% | 32.7% | 20.1% | 9.3% | 0.0% | -8.1% | -15.3% | -21.6% | -21.6% |
| | 2.0% | 60.9% | 43.5% | 29.0% | 16.7% | 6.2% | -2.8% | -10.7% | -17.7% | -17.7% |
| | 2.5% | 77.2% | 56.4% | 39.4% | 25.3% | 13.4% | 3.2% | -5.6% | -13.3% | -13.3% |
| | 3.0% | 97.5% | 72.2% | 52.0% | 35.5% | 21.8% | 10.2% | 0.3% | -8.2% | -8.2% |

Table 5: Case Study: Sensitivity Analysis WACC, Perpetual growth rate

The table clearly shows that even slight changes in the WACC or in the perpetual growth rate, which might not even be significant from an economist's perspective, will largely offset the determined fair share price from the base case scenario. For example increasing the WACC by 100bp and simultaneously decreasing the perpetual growth rate by 50bp will shrink the calculated fair stock price by more than 19%. Since it is very difficult to estimate the perpetual growth rate or the cost of capital with an exactness of just a few base points, the determined fair share price can only be seen as guidance, but not as an absolutely exact value.

The sensitivity to changes in the WACC can be expressed as the first derivative of the company value in respect to the discount rate, similar to the concept of bond duration. The formula below shows the approximate change in the company value when modifying the WACC.[15]

$$\frac{dV}{dr} = \frac{1}{1+r} \sum_{t=0}^{n} \frac{-t * FCF_t}{(1+r)^t}$$

The next step in the sensitivity analysis is to assess whether changes in the perpetual growth rate or in the growth rate for the predicted period (Sales CAGR) have a higher impact on the share price. Since both growth rates affect the nominal value free cash flow, the result of the analysis should be helpful to understand the importance that the terminal value has on the DCF analysis since all other factors are kept fixated. If modifying the perpetual growth rate leads to larger changes than modifying the sales CAGR for the scenario period, the terminal value would be of significantly higher importance than the scenario predictions for the first years.

---

[15]Due to convexity however, this approximation should only be used in the case of small changes in the discount rate.

| | | Perpetual growth rate (%) | | | | | | | | |
|---|---|---|---|---|---|---|---|---|---|---|
| | | 0.50% | 0.75% | 1.00% | 1.25% | 1.50% | 1.75% | 2.00% | 2.25% | 2.50% |
| Sales CAGR (%) | 6.75% | -14.0% | -11.8% | -9.4% | -6.9% | -4.3% | -1.4% | 1.6% | 4.9% | 8.5% |
| | 7.00% | -12.8% | -10.5% | -8.1% | -5.6% | -2.9% | 0.0% | 3.2% | 6.5% | 10.1% |
| | 7.25% | -11.5% | -9.2% | -6.8% | -4.2% | -1.4% | 1.5% | 4.7% | 8.1% | 11.8% |
| | 7.50% | -10.3% | -7.9% | -5.5% | -2.8% | 0.0% | 3.0% | 6.2% | 9.7% | 13.4% |
| | 7.75% | -9.0% | -6.6% | -4.1% | -1.4% | 1.5% | 4.5% | 7.8% | 11.3% | 15.1% |
| | 8.00% | -7.7% | -5.3% | -2.7% | 0.0% | 2.9% | 6.1% | 9.4% | 13.0% | 16.9% |
| | 8.25% | -6.4% | -3.9% | -1.3% | 1.5% | 4.5% | 7.6% | 11.0% | 14.7% | 18.6% |

Table 6: Case Study: Sensitivity analysis perpetual growth rate, sales CAGR

As expected, changes in the perpetual growth rate have a higher impact than changes in the sales CAGR have. For example an increase in the perpetual growth rate by 25bp result in a 3% higher share price, whereas a change by the same amount in the sales CAGR will only drive the fair share price up by 1.5%. Looking at this result, the importance of the terminal value becomes evident again. It underlines the fact that the TV includes all cash flows from the end of the scenario period up to infinity compared to just a few years in the scenario period. Therefore the TV, together with its underlying assumptions, is the most important and influential part of the whole discounted cash flow analysis. As mentioned before it is very easy to slightly adjust the assumptions that influence the TV, without having to justify these changes since they are very small. However, these small adjustments will significantly change the TV and therefore the value of the whole company.

## 5    Conclusion

The sensitivity analysis has shown that the DCF method is very vulnerable to changes in the underlying assumptions. Only marginally changes in the perpetual growth rate will lead to huge variances in the terminal value. Since the terminal value accounts for a large portion of the company's value, this is of big significance for the validity of the DCF method.

It is very easy to manipulate the DCF analysis to result in the value that you want it to result in by adjusting the inputs. This is even possible without making changes that would be significant from an economist's point of perspective, e.g. a change in the perpetual growth rate or in the WACC by just a few base points. Analysts or business professionals have no tools to estimate the input factors with that kind of exactness.

However, the DCF analysis is a great tool to analyze what assumptions and conditions have to be fulfilled in order to reach a certain company value. This is especially helpful in the case of capital budgeting and in the creation of feasibility plans.

The company valuation using discounted cash flows is a valid method to assess the company's value if special precaution is put on the validity of the underlying assumptions. As with all other financial models, the validity of the DCF method almost completely depends on the quality and validity of the data that is used as input. If used wisely, the discounted cash flow valuation is a powerful tool to evaluate the values of a variety of assets and also to analyze the effects that different economic scenarios have on a company's value.

The range of reasonable rates for discount factor and perpetual growth rate depends on each specific firm, its business situation and many more variables. In general you can say that the more risky a firm is, the higher its capital costs (WACC) are. The perpetual growth rate should be the same for all industries, since according to the arbitrage theory in the long run all companies and industries will grow by the same rate.

I conclude that using the DCF method in combination with other methods, like the trading comparables or precedent transaction analysis, is an effective approach to obtain a realistic range of appropriate company values. This combination technique is indeed the method that most companies and investment banks use today. When using several valuation techniques, their individual shortfalls are eliminated and the ultimate goal in the field of company valuation can be reached: determining a fair and valid company value.

## Appendix

Table 1: Long term credit rating scales

| Rating Agency | Moody's | Standard & Poor's (S&P) | Fitch |
|---|---|---|---|
| | Aaa | AAA | AAA |
| | Aa1 | AA+ | AA+ |
| | Aa2 | AA | AA |
| | Aa3 | AA- | AA- |
| **Investment grade debt** | A1 | A+ | A+ |
| | A2 | A | A |
| | A3 | A- | A- |
| | Baa1 | BBB+ | BBB+ |
| | Baa2 | BBB | BBB |
| | Baa3 | BBB- | BBB- |
| | Ba1 | BB+ | BB+ |
| | Ba2 | BB | BB |
| | Ba3 | BB- | BB- |
| | B1 | B+ | B+ |
| **Non-investment grade debt** | B2 | B | B |
| | B3 | B- | B- |
| | Caa1 | CCC+ | CCC+ |
| | Caa2 | CCC | CCC |
| | Caa3 | CCC- | CCC- |
| | Ca | CC | CC |
| | C | C | C |
| **Default grade debt** | C | D | D |

Table 2: Trading comparables analysis

| Company | EV/Sales | | | EV / EBITDA | | | EV / EBIT | | | Eq. V. / Net income | | |
|---|---|---|---|---|---|---|---|---|---|---|---|---|
| | 2008e | 2009e | 2010e | 2008e | 2009e | 2010e | 2008e | 2009e | 2010e | 2008e | 2009e | 2010e |
| Sixt | 1.0x | 1.0x | 1.0x | 3.2x | 3.1x | 2.9x | 9.2x | 8.9x | 8.4x | 7.3x | 7.1x | 6.7x |
| Avis Europe | 0.8x | 0.8x | 0.8x | 2.4x | 2.3x | 2.1x | 9.9x | 8.9x | 8.2x | 6.6x | 5.3x | 4.6x |
| D'ieteren | 0.5x | 0.5x | 0.4x | 4.0x | 3.7x | 3.5x | 8.2x | 7.6x | 6.8x | 6.9x | 5.9x | 5.0x |
| Hertz | 1.7x | 1.6x | 1.6x | 9.5x | 8.7x | 8.2x | 10.9x | 10.3x | 9.9x | 8.4x | 7.0x | 6.0x |
| Dollar Thrifty | 1.5x | 1.4x | 1.4x | 29.5x | 26.0x | 28.3x | | | | 12.3x | 8.0x | 8.4x |
| Penske | 0.3x | 0.3x | 0.2x | 10.2x | 9.3x | 8.0x | 14.4x | 12.4x | | 10.7x | 9.6x | 9.3x |
| Amerco | | | | | | | | | | | | |
| *Mean* | *1.0x* | *0.9x* | *0.9x* | *9.8x* | *8.8x* | *8.8x* | *10.5x* | *9.6x* | *8.3x* | *8.7x* | *7.2x* | *6.7x* |
| *Median* | *0.9x* | *0.9x* | *0.9x* | *6.7x* | *6.2x* | *5.8x* | *9.9x* | *8.9x* | *8.3x* | *7.8x* | *7.1x* | *6.3x* |

Table 3: Transaction multiple analysis

| Target | Acquirer | Date | EV (€m) | EV / SALES | EV / EBITDA | EV / EBIT | EqV / Net Income |
|---|---|---|---|---|---|---|---|
| Vanguard Car Rental EMEA | Europcar International | 13/11/2006 | 670.00 | 1.70x | 6.34x | 23.92x | n.m. |
| Keddy Car | Europcar International | 30/06/2006 | 0.00 | | | | |
| Europcar International | Eurazeo SA | 03.09.2006 | 3083.00 | 2.41x | | | |
| Hertz Group (Canada) | FirstGroup plc | 20/12/2000 | 18.07 | 1.22x | | | |
| Laidlaw International | FirstGroup plc | 02.09.2007 | 2701.76 | 1.11x | 7.43x | 13.84x | 22.10x |
| Cognisa Transportation | First Transit, Inc | 01.05.2007 | 11.87 | | | | |
| SKE Support Services | FirstGroup plc | 13/09/2004 | 22.85 | 0.38x | | | |
| Aircoach | FirstGroup plc | 11.01.2003 | 16.99 | | | | |
| GB Railways Group | FirstGroup plc | 16/07/2003 | 44.51 | 0.34x | | 29.67x | 55.64x | 88.99x |
| Coach USA | Kohlberg & Company LLC | 06.06.2003 | 130.99 | 0.72x | | | |
| Verona Bus Service | FirstGroup plc | 08.01.2001 | 6.51 | 1.00x | 3.81x | 7.15x | |
| Avis Greece | Piraeus Bank SA | 05.02.2007 | 215.50 | 2.65x | | | |
| Avis French | Avis Europe plc | 02.03.2003 | 8.50 | 0.43x | | | |
| Budget International | Avis Europe plc | 23/01/2003 | 37.28 | | | | |
| SAISC | Avis Europe plc | 31/01/2002 | 25.58 | | | | |
| 3 Arrows | Avis Europe plc | 12.10.1998 | 57.09 | | | | |
| Fraikin SA | CVC Capital | 12.08.2006 | 1350.00 | 2.21x | | 18.10x | |
| **Average** | | | | **1.29x** | **11.81x** | **23.73x** | **55.54x** |
| **Median** | | | | **1.11x** | **6.89x** | **18.10x** | **55.54x** |

Table 4: Case Study: Calculation of the enterprise value

| Period | 2008E | 2009E | 2010E | 2011E | 2012E | 2013E | TV |
|---|---|---|---|---|---|---|---|
| FCFF | 4,284 | 4,405 | 4,866 | 5,409 | 6,148 | 6,212 | - |
| NPV | 3,930 | 3,708 | 3,758 | 3,832 | 3,996 | 3,704 | 44,923 |
| EV | 67,850 | | | | | | |

Table 5: Case Study: Sensitivity Analysis WACC, perpetual growth rate

| | | WACC (%) | | | | | | | | |
|---|---|---|---|---|---|---|---|---|---|---|
| | | 7.0% | 7.5% | 8.0% | 8.5% | 9.0% | 9.5% | 10.0% | 10.5% | 11.0% |
| **Perpetual growth rate (%)** | 0.0% | 19.2% | 9.0% | 0.2% | -7.6% | -14.5% | -20.7% | -26.2% | -31.2% | -31.2% |
| | 0.5% | 27.2% | 15.8% | 5.9% | -2.7% | -10.3% | -17.0% | -23.0% | -28.3% | -28.3% |
| | 1.0% | 36.6% | 23.6% | 12.5% | 2.9% | -5.4% | -12.8% | -19.3% | -25.1% | -25.1% |
| | 1.5% | 47.6% | 32.7% | 20.1% | 9.3% | 0.0% | -8.1% | -15.3% | -21.6% | -21.6% |
| | 2.0% | 60.9% | 43.5% | 29.0% | 16.7% | 6.2% | -2.8% | -10.7% | -17.7% | -17.7% |
| | 2.5% | 77.2% | 56.4% | 39.4% | 25.3% | 13.4% | 3.2% | -5.6% | -13.3% | -13.3% |
| | 3.0% | 97.5% | 72.2% | 52.0% | 35.5% | 21.8% | 10.2% | 0.3% | -8.2% | -8.2% |

Table 6: Case Study: Sensitivity analysis perpetual growth rate, sales CAGR

| | | Perpetual growth rate (%) | | | | | | | | |
|---|---|---|---|---|---|---|---|---|---|---|
| | | 0.50% | 0.75% | 1.00% | 1.25% | 1.50% | 1.75% | 2.00% | 2.25% | 2.50% |
| **Sales CAGR (%)** | 6.75% | -14.0% | -11.8% | -9.4% | -6.9% | -4.3% | -1.4% | 1.6% | 4.9% | 8.5% |
| | 7.00% | -12.8% | -10.5% | -8.1% | -5.6% | -2.9% | 0.0% | 3.2% | 6.5% | 10.1% |
| | 7.25% | -11.5% | -9.2% | -6.8% | -4.2% | -1.4% | 1.5% | 4.7% | 8.1% | 11.8% |
| | 7.50% | -10.3% | -7.9% | -5.5% | -2.8% | 0.0% | 3.0% | 6.2% | 9.7% | 13.4% |
| | 7.75% | -9.0% | -6.6% | -4.1% | -1.4% | 1.5% | 4.5% | 7.8% | 11.3% | 15.1% |
| | 8.00% | -7.7% | -5.3% | -2.7% | 0.0% | 2.9% | 6.1% | 9.4% | 13.0% | 16.9% |
| | 8.25% | -6.4% | -3.9% | -1.3% | 1.5% | 4.5% | 7.6% | 11.0% | 14.7% | 18.6% |

Table 7: Case Study: Income statement estimates

| Period | 2008E | 2009E | 2010E | 2011E | 2012E | 2013E | 2014E |
|---|---|---|---|---|---|---|---|
| **Sales** | 64,702.10 | 65,388.80 | 67,645.50 | 71,390.10 | 74,631.10 | 76,870.00 | 86,517.90 |
| **EBIT Margin** | 12.2% | 11.2% | 11.9% | 13.2% | 14.0% | 13.0% | 11.0% |
| | | | | | | | |
| **EBIT** | 7,893.66 | 7,323.55 | 8,049.81 | 9,423.49 | 10,448.35 | 9,993.10 | 9,516.97 |
| **Taxes** | (2,368.10) | (2,197.06) | (2,414.94) | (2,827.05) | (3,134.51) | (2,997.93) | (2,855.09) |
| | | | | | | | |
| **NOPLAT** | 5,525.56 | 5,126.48 | 5,634.87 | 6,596.45 | 7,313.85 | 6,995.17 | 6,661.88 |
| | | | | | | | |
| **D&A** | 2,700.90 | 2,740.90 | 2,779.20 | 2,829.50 | 2,902.30 | 2,989.40 | 3,431.80 |
| **Increase in NWC** | (1,031.20) | (519.80) | (503.60) | (804.20) | (709.70) | (783.60) | (593.80) |
| **Capex** | (2,911.60) | (2,942.50) | (3,044.00) | (3,212.60) | (3,358.40) | (2,989.40) | (3,431.80) |
| | | | | | | | |
| **FCFF** | 4,283.66 | 4,405.08 | 4,866.47 | 5,409.15 | 6,148.05 | 6,211.57 | 6,068.08 |

Table 8: Case Study: Liabilities structure

| | | |
|---|---|---|
| **Shareholders Equity** | | 20,097.90 |
| **Financial Debt** | | 10,100.70 |
| | Long Term | 6,953.00 |
| | Short Term | 3,147.70 |
| | | |
| **Leverage** | | 0.33 |

Table 9: Case Study: WACC calculation

| | |
|---|---|
| **Cost of Equity (%)** | |
| Risk free rate (%) | 4.3% |
| Unlevered Beta | 0.9 |
| Levered Beta | 1.2 |
| Market return (%) | 9.3% |
| | |
| CAPM required RoE | 10.3% |
| | |
| **Cost of Debt (%)** | |
| Average Credit Spread (%) | 5.0% |
| Cost of Debt before taxes | 9.3% |
| CoD adjusted for tax | 6.5% |
| | |
| **WACC** | **9.0%** |

Table 10: Case Study: Terminal Value calculation

| | |
|---|---|
| FCFF in terminal period | 6,068.08 |
| Perpetual growth rate (%) | 1.5% |
| WACC (%) | 9.0% |
| Terminal Value | 81,731.65 |
| NPV of TV | 44,607.47 |

Table 11: Case Study: DCF valuation

| Period | 2008E | 2009E | 2010E | 2011E | 2012E | 2013E | TV |
|---|---|---|---|---|---|---|---|
| FCFF (EURm) | 4,283.66 | 4,405.08 | 4,866.47 | 5,409.15 | 6,148.05 | 6,211.57 | - |
| NPV (EURm) | 3,929.96 | 3,707.67 | 3,757.81 | 3,831.97 | 3,995.81 | 3,703.76 | 44,923.18 |
| EV (EURm) | 67,850.16 | | | | | | |
| Net debt (EURm) | (11,547.00) | | | | | | |
| Minorities (EURm) | (971.20) | | | | | | |
| Eq.V. (EURm) | 55,331.96 | | | | | | |
| No. Of shares (m) | 946 | | | | | | |
| Fair share price | 58.49 | | | | | | |

Figure 1: LIBOR credit spread (in bp)

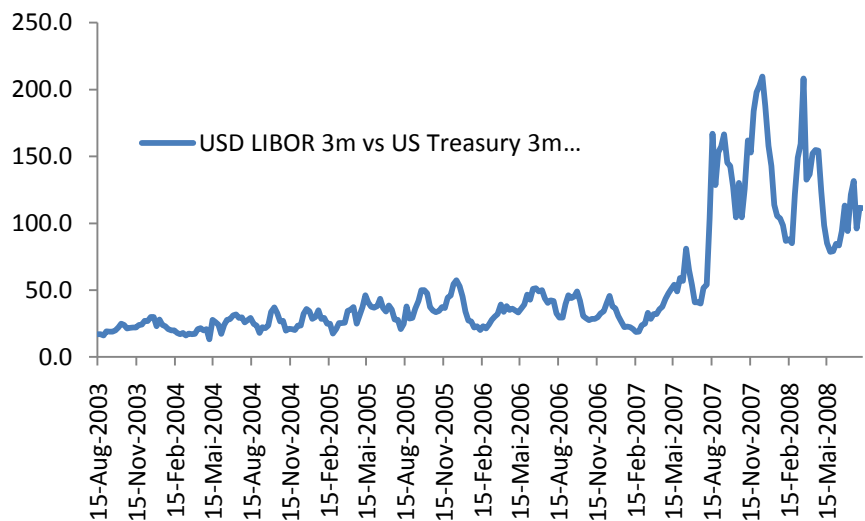